\documentstyle[twoside]{article}

\catcode`\@=11
\long\def\@makefntext#1{
\protect\noindent \hbox to 3.2pt {\hskip-.9pt
$^{{\eightrm\@thefnmark}}$\hfil}#1\hfill}               

\def\thefootnote{\fnsymbol{footnote}}
\def\@makefnmark{\hbox to 0pt{$^{\@thefnmark}$\hss}}    

\def\ps@myheadings{\let\@mkboth\@gobbletwo
\def\@oddhead{\hbox{}
\rightmark\hfil\eightrm\thepage}
\def\@oddfoot{}\def\@evenhead{\eightrm\thepage\hfil
\leftmark\hbox{}}\def\@evenfoot{}
\def\sectionmark##1{}\def\subsectionmark##1{}}

\oddsidemargin=\evensidemargin
\renewcommand{\thefootnote}{\fnsymbol{footnote}}

\newcounter{sectionc}\newcounter{subsectionc}\newcounter{subsubsectionc}
\renewcommand{\section}[1] {\vspace{12pt}\addtocounter{sectionc}{1}
\setcounter{subsectionc}{0}\setcounter{subsubsectionc}{0}\noindent
        {\tenbf\thesectionc. #1}\par\vspace{5pt}}
\renewcommand{\subsection}[1] {\vspace{12pt}\addtocounter{subsectionc}{1}
        \setcounter{subsubsectionc}{0}\noindent
        {\bf\thesectionc.\thesubsectionc. {\kern1pt \bfit #1}}\par\vspace{5pt}}
\renewcommand{\subsubsection}[1] {\vspace{12pt}\addtocounter{subsubsectionc}{1}
        \noindent{\tenrm\thesectionc.\thesubsectionc.\thesubsubsectionc.
        {\kern1pt \tenit #1}}\par\vspace{5pt}}
\newcommand{\nonumsection}[1] {\vspace{12pt}\noindent{\tenbf #1}
        \par\vspace{5pt}}

\newcounter{appendixc}
\newcounter{subappendixc}[appendixc]
\newcounter{subsubappendixc}[subappendixc]
\renewcommand{\thesubappendixc}{\Alph{appendixc}.\arabic{subappendixc}}
\renewcommand{\thesubsubappendixc}
        {\Alph{appendixc}.\arabic{subappendixc}.\arabic{subsubappendixc}}

\renewcommand{\appendix}[1] {\vspace{12pt}
        \refstepcounter{appendixc}
        \setcounter{figure}{0}
        \setcounter{table}{0}
        \setcounter{lemma}{0}
        \setcounter{theorem}{0}
        \setcounter{corollary}{0}
        \setcounter{definition}{0}
        \setcounter{equation}{0}
        \renewcommand{\thefigure}{\Alph{appendixc}.\arabic{figure}}
        \renewcommand{\thetable}{\Alph{appendixc}.\arabic{table}}
        \renewcommand{\theappendixc}{\Alph{appendixc}}
        \renewcommand{\thelemma}{\Alph{appendixc}.\arabic{lemma}}
        \renewcommand{\thetheorem}{\Alph{appendixc}.\arabic{theorem}}
        \renewcommand{\thedefinition}{\Alph{appendixc}.\arabic{definition}}
        \renewcommand{\thecorollary}{\Alph{appendixc}.\arabic{corollary}}
        \renewcommand{\theequation}{\Alph{appendixc}.\arabic{equation}}
        \noindent{\tenbf Appendix \theappendixc #1}\par\vspace{5pt}}
\newcommand{\subappendix}[1] {\vspace{12pt}
        \refstepcounter{subappendixc}
        \noindent{\bf Appendix \thesubappendixc. {\kern1pt \bfit #1}}
        \par\vspace{5pt}}
\newcommand{\subsubappendix}[1] {\vspace{12pt}
        \refstepcounter{subsubappendixc}
        \noindent{\rm Appendix \thesubsubappendixc. {\kern1pt \tenit #1}}
        \par\vspace{5pt}}

\newcommand{\textlineskip}{\baselineskip=13pt}
\newcommand{\smalllineskip}{\baselineskip=10pt}

\def\eightcirc{
\begin{picture}(0,0)
\put(4.4,1.8){\circle{6.5}}
\end{picture}}
\def\eightcopyright{\eightcirc\kern2.7pt\hbox{\eightrm c}}

\newcommand{\copyrightheading}[1]
        {\vspace*{-2.5cm}\smalllineskip{\flushleft
        {\footnotesize $\eightcopyright$\, World Scientific Publishing
         Company}\\
         }}

\def\abstracts#1#2#3{{
        \centering{\begin{minipage}{4.5in}\baselineskip=10pt\footnotesize
        \parindent=0pt #1\par
        \parindent=15pt #2\par
        \parindent=15pt #3
        \end{minipage}}\par}}

\newcommand{\bibit}{\nineit}
\newcommand{\bibbf}{\ninebf}
\renewenvironment{thebibliography}[1]
        {\frenchspacing
         \ninerm\baselineskip=11pt
         \begin{list}{\arabic{enumi}.}
        {\usecounter{enumi}\setlength{\parsep}{0pt}
         \setlength{\leftmargin 12.7pt}{\rightmargin 0pt} 
         \setlength{\itemsep}{0pt} \settowidth
        {\labelwidth}{#1.}\sloppy}}{\end{list}}

\newcounter{itemlistc}
\newcounter{romanlistc}
\newcounter{alphlistc}
\newcounter{arabiclistc}
\newenvironment{itemlist}
        {\setcounter{itemlistc}{0}
         \begin{list}{$\bullet$}
        {\usecounter{itemlistc}
         \setlength{\parsep}{0pt}
         \setlength{\itemsep}{0pt}}}{\end{list}}

\newcommand{\fcaption}[1]{
        \refstepcounter{figure}
        \setbox\@tempboxa = \hbox{\footnotesize Fig.~\thefigure. #1}
        \ifdim \wd\@tempboxa > 5in
           {\begin{center}
        \parbox{5in}{\footnotesize\smalllineskip Fig.~\thefigure. #1}
            \end{center}}
        \else
             {\begin{center}
             {\footnotesize Fig.~\thefigure. #1}
              \end{center}}
        \fi}

\newcommand{\tcaption}[1]{
        \refstepcounter{table}
        \setbox\@tempboxa = \hbox{\footnotesize Table~\thetable. #1}
        \ifdim \wd\@tempboxa > 5in
           {\begin{center}
        \parbox{5in}{\footnotesize\smalllineskip Table~\thetable. #1}
            \end{center}}
        \else
             {\begin{center}
             {\footnotesize Table~\thetable. #1}
              \end{center}}
        \fi}

\def\@citex[#1]#2{\if@filesw\immediate\write\@auxout
        {\string\citation{#2}}\fi
\def\@citea{}\@cite{\@for\@citeb:=#2\do
        {\@citea\def\@citea{,}\@ifundefined
        {b@\@citeb}{{\bf ?}\@warning
        {Citation `\@citeb' on page \thepage \space undefined}}
        {\csname b@\@citeb\endcsname}}}{#1}}

\newif\if@cghi
\def\cite{\@cghitrue\@ifnextchar [{\@tempswatrue
        \@citex}{\@tempswafalse\@citex[]}}
\def\citelow{\@cghifalse\@ifnextchar [{\@tempswatrue
        \@citex}{\@tempswafalse\@citex[]}}
\def\@cite#1#2{{$\null^{#1}$\if@tempswa\typeout
        {IJCGA warning: optional citation argument
        ignored: `#2'} \fi}}

\def\pmb#1{\setbox0=\hbox{#1}
        \kern-.025em\copy0\kern-\wd0
        \kern.05em\copy0\kern-\wd0
        \kern-.025em\raise.0433em\box0}

\def\fnt#1#2{\footnotetext{\kern-.3em
        {$^{\mbox{\scriptsize #1}}$}{#2}}}

\def\runninghead#1#2{\pagestyle{myheadings}
\markboth{{\protect\footnotesize\it{\quad #1}}\hfill}
{\hfill{\protect\footnotesize\it{#2\quad}}}}
\font\tenrm=cmr10
\font\tenit=cmti10
\font\tenbf=cmbx10
\font\bfit=cmbxti10 at 10pt
\font\ninerm=cmr9
\font\nineit=cmti9
\font\ninebf=cmbx9
\font\eightrm=cmr8

\def\qed{\hbox{${\vcenter{\vbox{                        
   \hrule height 0.4pt\hbox{\vrule width 0.4pt height 6pt
   \kern5pt\vrule width 0.4pt}\hrule height 0.4pt}}}$}}

\renewcommand{\thefootnote}{\fnsymbol{footnote}}        

\def\bsc{{\sc a\kern-6.4pt\sc a\kern-6.4pt\sc a}}       
\def\bflatex{\bf L\kern-.30em\raise.3ex\hbox{\bsc}\kern-.14em
T\kern-.1667em\lower.7ex\hbox{E}\kern-.125em X}

\newcommand{\tr}{{\rm\;tr}}
\newcommand{\Tr}{{\rm\;Tr}}
\newcommand{\comm}[2]{\left[#1,#2\right]}
\newcommand{\DD}{{\cal D}}

\renewcommand{\d}{{\rm d}}
\newcommand{\re}{\Re{\rm e}}
\renewcommand{\i}{{\rm i}}
\newlength{\newdisplayskip}
\begin{document}

%
%
\runninghead{A Basis for Invariants in Non--Abelian Gauge Theories}
{A Basis for Invariants in Non--Abelian Gauge Theories}

\normalsize\textlineskip
\thispagestyle{empty}
\setcounter{page}{1}
\begin{flushright}
  {\large hep-th/9508031}
\end{flushright}
\vspace*{0.8truein}

%
%
\centerline{\bf A BASIS FOR INVARIANTS IN NON--ABELIAN GAUGE
  THEORIES\footnote{Talk presented at the AIHENP--95 workshop, Pisa
    (Italy), April 1995}}
 \vspace*{0.37truein}
\centerline{\footnotesize UWE M\"ULLER\footnote{E--mail address:
    umueller@convex.ifh.de}}
\vspace*{0.015truein}
\centerline{\footnotesize\it Theory Group, DESY--IfH Zeuthen,
  Platanenallee 6}
\baselineskip=10pt
\centerline{\footnotesize\it D--15738 Zeuthen, Germany}

\setcounter{footnote}{0}
\renewcommand{\thefootnote}{\rm\alph{footnote}}
\vspace*{0.21truein}
\abstracts{
An algorithm is described to convert Lorentz and gauge invariant
expressions in
non--Abelian gauge theories with matter into a standard form,
consisting of a linear combination of basis invariants. This
algorithm is needed for computer calculations of effective
actions. The defining
properties of the basis invariants are reported. The number of basis
invariants up to mass dimension 16 are presented.}
{}{}

\vspace*{1pt}\textlineskip      
\section{Introduction}          
\noindent
\noindent
Effective actions of gauge theories are
space--time integrals over gauge and Lorentz invariant
expressions. From the mathematical point of view, they are, up to some
factors, functional traces of heat kernel coefficients, known as
Schwinger--DeWitt,\cite{SD} Gilkey--Seeley,\cite{GS} or Hadamard
coefficients.\cite{H} In flat
space-time, these coefficients are polynomials
constructed from a matrix potential and from the gauge field strength
tensor by multiplication, gauge covariant differentiation, and
contraction of Lorentz indices. Due to Bianchi identities and the
product rule for covariant derivatives, the form of the coefficients is not
unique. Furthermore, the physically interesting functional trace
of the coefficients allows cyclic exchanges of matrix factors and
integration by parts.

New methods of computing effective actions, such as the
string--inspired world line path integral formalism,\cite{worldline
  formalism,fliegner} but also the
implementation of established calculation algorithms on
computers\cite{lanyov} enable the extension of known
results to higher order in the inverse mass expansion.
To manage the corresponding increasing number of terms
and to compare results of different
methods,\cite{methods,results} a standard
basis of invariants is needed, in terms of
which all results can be expressed. An algorithm should be
provided to convert a Lorentz scalar given in a non--standard form
into terms of the basis.
For gravitational invariants, constructed from the Riemann and
the metric tensor, such normal forms were presented up to order eight in
the mass dimension by Fulling et al.\cite{Fulling et al.}
In the general case with matter, gauge fields, and gravity,
basis sets of non--local invariants up to third order in the
curvature were constructed. They are used in the
expansion of effective actions in terms of Barvinsky--Vilkovisky
form factors.\cite{BV}

This contribution analyzes the formal structure of invariant monomials
in non--Abelian gauge theories with matter in flat space--time. Step
by step, the operations applicable to invariants are used to
convert them into a fixed form. Thus, a basis of invariants is
specified, and simultaneously, a procedure to expand an arbitrary
given Lorentz invariant expression in terms of the basis is
obtained. The proof of the basis property of the specified set of
invariants will be published elsewhere.\cite{proof}

\section{Notations}
\label{graph.not}
\noindent
Notations are introduced on the basis
of a concrete example. Let us consider a gauged scalar field
theory described by the massive complex field $\phi^a$ and the
Hermitian matrix valued gauge field $A_\mu^{ab}$. The gauge
covariant derivative in the fundamental representation is
${\cal D}_\mu^{ab}=\delta^{ab}\partial_\mu-\i A_\mu^{ab}$. The
coupling constant is contained in the gauge field.
Integrating the quantum fluctuations of the field $\phi^a$ in the
given backgrounds $\varphi^a$ and $A_\mu^a$, we obtain, in a first
approximation, the one--loop effective action
$\Gamma^{(1)}[\varphi,A]$ which can be expanded in gauge invariant
terms\cite{methods}
\vspace*{-0.2cm}
\begin{equation}\label{generic result}
  \Gamma^{(1)}[\varphi,A]=\Tr\ln\left(-\DD^2+V+m^2\right)=
  \int\d^dx\sum_i\frac{C_i}{m^{\mu_i-d}}\tr\left(I_i\left(F,V\right)\right).
\vspace*{-0.3cm}
\end{equation}
$V$ is a matrix potential originating from the matter fields.
The $C_i$ are complex numbers and $I_i\left(F,V\right)$ matrix valued
Lorentz scalars composed of the potential $V$,
the field strength tensor
\begin{math}
  F_{\mu\nu}^{ab}=\i\comm{\DD_\mu}{\DD_\nu}^{ab}=
  \partial_\mu A_\nu^{ab}-\partial_\nu A_\mu^{ab}-
  \i\comm{A_\mu}{A_\nu}^{ab},
\end{math}
and the gauge covariant derivative in the adjoint representation
\begin{math}
  D_\mu =\comm{\DD_\mu}{.}=\partial_\mu -\i \comm{A_\mu}{.}.
\end{math}
$D_\mu$ acts on the matrix potential and on the field strength
tensor. $d$ is the dimension of space--time. $\mu_i$ is the mass
dimension of the scalar $I_i(F,V)$ according to the
mass dimensions of its constituents $[V]=2$,
$\left[F_{\mu\nu}\right]=2$, and
$\left[\DD_\mu\right]=\left[D_\mu\right]=1$.

The form (\ref{generic result}) is not unique due to
several equalities, namely the product rule for
covariant derivatives, integration by parts, cyclic permutations,
the Bianchi identity, the antisymmetry of the field strength tensor,
and the exchange of derivatives:
\begin{displaymath}
\refstepcounter{equation}
\label{manipulations}
\begin{array}{c@{\quad}cr}
   D_\mu(XY)=D_\mu XY+XD_\mu Y,&
   \int\d x\tr\left(D_\mu X_\mu Y\right)
    =-\int\d x\tr\left(X_\mu D_\mu Y\right)\!,&
    \makebox[0.8cm][r]{(\theequation\rm a,b)\hspace{-0.2cm}}\\
    \tr(XY\ldots Z)=\tr(Y\ldots ZX), &
    D_\mu F_{\kappa\lambda}=D_\kappa F_{\mu\lambda}+D_\lambda
    F_{\kappa\mu},&
    \makebox[0.45cm][r]{(\theequation\rm c,d)\hspace{-0.2cm}}\\
    F_{\mu\nu}=-F_{\nu\mu},&
    D_\mu D_\nu X=D_\nu D_\mu X-\i\comm{F_{\mu\nu}}{X}.&
    \makebox[0.45cm][r]{(\theequation\rm e,f)\hspace{-0.2cm}}
\end{array}
\end{displaymath}

Let us call a $V$, an $F$, or covariant derivatives of them
a {\em simple factor\/}, i.e.
\vspace*{-0.1cm}
\begin{equation}\label{factors}
  (\mbox{simple factor})\in\{
    V,\;
    F_{\kappa\lambda},\;
    D_{\mu_1}D_{\mu_2}\ldots D_{\mu_n}V,\;
    D_{\mu_1}D_{\mu_2}\ldots D_{\mu_n}F_{\kappa\lambda}\}.\vspace*{-0.1cm}
\end{equation}
Simple factors containing the matrix potential are called
{\em $V$--factors}, the others {\em $F$--factors}. With the
product rule (\ref{manipulations}a), expression (\ref{generic result})
can be converted into a form where the invariants $I_i(F,V)$ are
monomials, i.e. products of simple factors.
Subsequently, the invariants are supposed to have this form.

If the gauge group representation is unitary, the
additional symmetries
\vspace*{-0.15cm}
\begin{equation}\label{symmetries}
  V^\dagger=V,\quad
  A_\mu^\dagger=A_\mu,\quad
  F_{\mu\nu}^\dagger=F_{\mu\nu},\quad
  \left(D_\mu X\right)^\dagger=D_\mu X\quad\mbox{if}\quad
  X^\dagger=X\vspace*{-0.15cm}
\end{equation}
hold. Consequently, simple factors are Hermitian. For simple factors
$X$, $Y$, and $Z$ this leads to
\vspace*{-0.4cm}
\begin{equation}\label{mirror transform}
  \overline{\tr\left(XYZ\ldots\right)}=
  \tr\left(\ldots Z^\dagger Y^\dagger X^\dagger\right)=
  \tr\left(\ldots ZYX\right).\vspace*{-0.15cm}
\end{equation}
Thus, an invariant monomial can be expressed by the complex conjugate
of its mirror image with identical factors, but in inverted order.
Therefore we call eq.~(\ref{mirror transform}) a mirror
transformation. In general, a monomial and its complex conjugate are
independent of each other, so that operation (\ref{mirror
  transform}) cannot
be used to reduce the number of terms in eq.~(\ref{generic result}).
However, Lagrangians are real. Hence, in an appropriate basis, an
arbitrary invariant monomial $I(F,V)$ and its mirror image have
complex conjugate coefficients $C$ and $\bar{C}$ so
that they add to $2\re\left(C\cdot I(F,V)\right)$. An\-other exception
occurs for real $\phi^a$ and imaginary $A_\mu^{ab}$.\footnote{This
  is the case for real orthogonal representations of the gauge
  group. Then $\i A_\mu^{ab}$ is real and antisymmetric in $a$ and $b$.}
Then $V$--factors are real and $F$--factors imaginary. In this
case, monomials and
their complex conjugates are not independent of each other and
eq.~(\ref{mirror transform}) reduces the number of terms in
eq.~(\ref{generic result}) indeed.

\section{The Basis}
\label{construction}
\vspace*{-0.4cm}
\subsection{The reduction algorithm}
\noindent
We start from an arbitrary Lorentz invariant given in the form
(\ref{generic result}). The product rule must be used whenever
derivatives of products are encountered. This may happen at each stage
of the algorithm. The
manipulations (\ref{manipulations}b--f, \ref{mirror transform}) must be
applied in the sequence of the following sub-subsections to obtain a
standard result. The rules given there do not
entirely fix all details of the algorithm. Therefore, the algorithm
can be executed in different ways, but
the results will be expressed by the same basis of
invariants and, hence, will be identical.
The procedure will require exchanges
of derivatives by eq.~(\ref{manipulations}f).
Since thereby additional invariants
with more $F$--factors and fewer derivatives are
produced, the algorithm starts with the invariants with the most
$F$--factors and descends to invariants with fewer
and fewer $F$--factors.

\subsubsection{Integration by parts}
\noindent
The indices in a Lorentz invariant monomial can be contracted
between different factors and within the same factor. We call the
latter self--contractions. They
always include a co\-var\-i\-ant derivative.
Therefore, {\it we apply integration by parts to covariant derivatives
  in self--contractions.} Thereby all self--contractions are
eliminated.

\subsubsection{The Bianchi identity}
\noindent
The Bianchi identity (\ref{manipulations}d) exchanges the index of
one derivative with the indices of $F_{\mu\nu}$ within an
$F$--factor. All other factors remain
unchanged. Therefore, we need a prescription that
specifies the derivatives which are candidates for applying the
Bianchi identity in the $F$--factor under consideration.
Let us consider the example
\vspace*{-0.3cm}
\begin{equation}\label{example sectors}
  \tr(
  \underbrace{
  \stackrel{\mbox{L}}{D_\mu}
  \stackrel{\mbox{R}}{D_\nu}
  \stackrel{\mbox{M}}{D_\rho}
  D_\sigma
  F_{\kappa\lambda}}_{\makebox[0cm]{\parbox{2.1cm}{\centering factor
      under\\consideration}}}
  \underbrace{\ldots X_\nu'\ldots}_{\parbox{1.5cm}{\centering right sector}}
  Y_{\sigma\kappa}
  \underbrace{\ldots X_\rho''\ldots}_{\parbox{1.5cm}
    {\centering middle sector}}Z_\lambda
  \underbrace{\rule{0cm}{1.85ex}
    \ldots X_\mu\ldots}_{\makebox[0cm][l]{\parbox{1.5cm}{\centering
        left sector}}})\quad.\vspace*{-0.1cm}
\end{equation}
The indices of $F_{\kappa\lambda}$ are contracted with the factors
$Y_{\sigma\kappa}$ and $Z_\lambda$, which divide the remaining factors
into three, possibly empty, sectors. We call them ``right sector'', ``middle
sector'', and ``left sector'', as indicated, because, due to cyclic
invariance (\ref{manipulations}c), the ``left sector'' is connected
with the left--hand side of the factor under consideration.

The derivatives of the factor under consideration are called
left (``L''), right (``R''), and middle (``M'') corresponding to the sector
they are contracted with. Not all derivatives are left, right, or
middle (e.g.~$D_\sigma$). The Bianchi identity (\ref{manipulations}d)
mixes all three kinds of derivatives. Therefore it can be used to
eliminate one kind of index in all factors of all monomials. Since
the middle sector is invariant under the mirror transformation (left and
right sectors are interchanged), {\it we apply the Bianchi identity to
  middle derivatives.} Each such application of the Bianchi identity
reduces the number of factors in the corresponding middle
sector. Thus, after finitely many steps, all middle derivatives are
eliminated.

Finally, we convert multiple contractions between factors
into a standard form by
\begin{eqnarray}\label{multiple1}\textstyle
  \ldots F_{\mu\nu}\ldots D_\mu D_\nu X\ldots&\Rightarrow&\textstyle
  -\frac{\i}{2}\ldots F_{\mu\nu}\ldots\comm{F_{\mu\nu}}{X}\ldots\\
  \label{multiple2}\textstyle
  \ldots F_{\mu\nu}\ldots D_\mu F_{\nu\kappa}\ldots&\Rightarrow&\textstyle
  \frac{1}{2}\ldots F_{\mu\nu}\ldots D_\kappa F_{\nu\mu}\ldots\\
  \label{multiple3}
  \ldots D_\mu F_{\nu\kappa}\ldots D_\nu F_{\mu\lambda}\ldots&\Rightarrow&
  \ldots D_\mu F_{\nu\kappa}\ldots D_\mu F_{\nu\lambda}\ldots+\nonumber\\
  &&\textstyle\hspace{1cm}+
  \frac{1}{2}\ldots D_\kappa F_{\nu\mu}\ldots D_\lambda F_{\mu\nu}\ldots\quad.
\end{eqnarray}
The first equality uses the antisymmetry
(\ref{manipulations}e) of the field strength tensor and the commutation
rule (\ref{manipulations}f). The second transformation relies on the
antisymmetry
(\ref{manipulations}e) together with the Bianchi identity
(\ref{manipulations}d). The third rule results by applying
the Bianchi identity (\ref{manipulations}d) to one of the factors
and subsequently using eq.~(\ref{multiple2}).

\subsubsection{The arrangement of factors}
\noindent
Cyclic factor permutations (\ref{manipulations}c) and, possibly, mirror
transformations (\ref{mirror transform}) can be used to identify
invariants. Applying eqs. (\ref{manipulations}c) and (\ref{mirror
  transform}) in all possible ways to a given invariant monomial,
we obtain a class of equivalent invariants. {\em We pick a
  representative of each equivalence class.} This may be done by
introducing an ordering relation in the equivalence classes
Then we pick the smallest (or greatest) invariant of each
equivalence class as the representative.

\subsubsection{The arrangement of indices}
\label{index convention}
\noindent
Derivatives and indices of $F$'s can be exchanged by means of
eqs.~(\ref{manipulations}f) and (\ref{manipulations}e) in all factors
of all invariant monomials. Let us consider a certain factor within an
invariant. It can be shifted completely to the left--hand side
by eq.~(\ref{manipulations}c), as a result of which the achieved
arrangement of factors is temporarily destroyed\footnote{The
  arrangement of the factors has to be restored after reordering
  the indices and is, in the end, not affected by this procedure.}
\hspace{0.1cm} (cf.\ example~(\ref{example sectors})).
{\it After this operation, we rearrange the derivatives and/or
  indices of the $F$ (if present) in the considered factor according
  to the contracted counter indices.\/}
In example~(\ref{example sectors}) $Y_{\sigma\kappa}$ is located left
of $Z_\lambda$. Thus the indices of $F_{\kappa\lambda}$ have the
correct order. The locations of $X_\nu'$, $Y_{\sigma\kappa}$,
$X_\rho''$, and $X_\mu$ define the correct order of the
derivatives to be $D_\nu D_\sigma D_\rho D_\mu$. Since the mirror
transformation inverts the ordering of the factors, it has to be applied
{\it before\/} rearranging the indices. Cyclic factor permutations
and the arrangement of indices do not interfere with each other.

\subsection{The defining properties of the basis}
\noindent
Pursuing the above algorithm,
we state the following properties of basis
invariants:\vspace*{-0.2cm}
\begin{itemlist}
  \item The invariants are products of simple factors.
  \item Indices are contracted only between different factors of an
    invariant monomial.
  \item There are no ``middle'' derivatives.
  \item In multiple contractions between factors, derivatives are
    contracted with de\-riv\-a\-tives and indices of $F$'s with indices of
    $F$'s (cf.\ eqs.~(\ref{multiple1} -- \ref{multiple3})) except for
    contractions of an index of an $F$ with a derivative where the
    other index of the $F$ is contracted with a third factor.
  \item The order of derivatives and of indices of the $F$'s is as
    described in sub-subsection 3.1.4.\vspace*{-0.2cm}
\end{itemlist}
These properties allow to count the basis invariants of a certain
mass dimension. Up to mass dimension 16, this was performed by a C
language program (table \ref{number of invariants}). Results of higher
dimension or divided by the number of $F$'s are available.
\begin{table}[htbp]
  \centering
  \tcaption{The number of basis invariants {\bf with} and {\it
      without} the mirror transformation. $v$ is the number of
    occurrences of the matrix potential $V$ in the
    invariants.}
  \label{number of invariants}
  \vspace*{1ex}\footnotesize\smalllineskip
  \raggedright
  \begin{tabular}{|c|c||rr||rr|rr|rr|}
    \hline
    &Mass dim.&\multicolumn{2}{c||}{Total}&
    \multicolumn{2}{c|}{$v=0$}&\multicolumn{2}{c|}{1}&
    \multicolumn{2}{c|}{2}\\
    \hline
    1&2&\bf 1&\it 1&\bf 0&\it 0&\bf 1&\it 1&&\\
    2&4&\bf 2&\it 2&\bf 1&\it 1&\bf 0&\it 0&\bf 1&\it 1\\
    3&6&\bf 5&\it 5&\bf 2&\it 2&\bf 1&\it 1&\bf 1&\it 1\\
    4&8&\bf 17&\it 18&\bf 7&\it 7&\bf 4&\it 5&\bf 4&\it 4\\
    5&10&\bf 79&\it 105&\bf 29&\it 36&\bf 24&\it 36&\bf 17&\it 23\\
    6&12&\bf 554&\it 902&\bf 196&\it 300&\bf 184&\it 329&\bf 119&\it 191\\
    7&14&\bf 5283&\it 9749&\bf 1788&\it 3218&\bf 1911&\it 3655&
    \bf 1096&\it 2020\\
    8&16&\bf 65346&\it 127072&\bf 21994&\it 42335&\bf 24252&\it 47844&
    \bf 13333&\it 25861\\
    \hline
  \end{tabular}\\[1ex]
  Table \ref{number of invariants}. (Continued)\\
  \begin{tabular}{|c|c||rr|rr|rr|rr|rr|rr|}
    \hline
    &Mass dim.&\multicolumn{2}{c|}{$v=3$}&\multicolumn{2}{c|}{4}&
    \multicolumn{2}{c|}{5}&\multicolumn{2}{c|}{6}&
    \multicolumn{2}{c|}{7}&\multicolumn{2}{c|}{8}\\
    \hline
    1&2&&&&&&&&&&&&\\
    2&4&&&&&&&&&&&&\\
    3&6&\bf 1&\it 1&&&&&&&&&&\\
    4&8&\bf 1&\it 1&\bf 1&\it 1&&&&&&&&\\
    5&10&\bf 6&\it 7&\bf 2&\it 2&\bf 1&\it 1&&&&&&\\
    6&12&\bf 39&\it 63&\bf 13&\it 16&\bf 2&\it 2&\bf 1&\it 1&&&&\\
    7&14&\bf 370&\it 670&\bf 96&\it 158&\bf 18&\it 24&\bf 3&\it 3&
    \bf 1&\it 1&&\\
    8&16&\bf 4452&\it 8638&\bf 1095&\it 2020&\bf 186&\it 329&
    \bf 30&\it 41&\bf 3&\it 3&\bf 1&\it 1\\
    \hline
  \end{tabular}\vspace*{-0.5cm}
\end{table}

\section{Conclusions and Outlook}
\noindent
A prescription for defining a standard basis set of invariants in
non--Abelian gauge theories was obtained. A reduction algorithm was
presented to
convert a given Lorentz scalar by partial integration, by the Bianchi
identity, and by cyclic invariance of the trace into a linear
combination of this basis set of invariants. The proof that this set
is a basis indeed, relies on a graphical representation of
invariants and is given elsewhere.\cite{proof}

For cases where, in addition, the mirror transformation reduces the
number of independent invariants, a general proof of the basis property
is still lacking. However at least up to mass dimension 16, it can be shown
by counting the invariants that the standard set remains a basis.

Another open problem is to take into account additional identities
which exist for particular choices of the gauge group representation.
\vspace*{-1mm}

\nonumsection{Acknowledgements}
\noindent
The author would like to thank C.~Schubert for indicating the need to
clarify this problem and D.~Fliegner for
discussions on the reduction algorithm. The proof--reading of the
manuscript by D.~Lehner and G.~Weigt is gratefully
acknowledged.\vspace*{-1mm}

\nonumsection{References}

\end{document}